\documentstyle[preprint,tighten,aps,epsf]{revtex} 

\newcommand{\bra}{\left\langle}
\newcommand{\ket}{\right\rangle}

\def\q{{\bf q}}
\def\p{{\bf p}}
\def\k{{\bf k}}
\def\n{{\bf n}}
\def\x{{\bf x}}
\def\v{{\bf v}}
\def\R{{\bf R}}

\def \be{\begin{equation}}
\def \ee{\end{equation}}

\begin{document}

\title{Classical and Quantum Dynamics in a Random Magnetic Field}
\author{A.V. Izyumov and B.D. Simons \\
{\small \em Cavendish Laboratory, Madingley Road, Cambridge, CB3 0HE, UK}}

\maketitle

\begin{abstract}

Using the supersymmetry approach, we study spectral statistical properties
of a two-dimensional quantum particle subject to a non-uniform magnetic field.
We focus mainly on the problem of regularisation of the field theory.
Our analysis begins with an 
investigation of the spectral properties of the purely classical evolution 
operator. We show that, although the kinetic equation is formally 
time-reversible, density relaxation is controlled by {\em irreversible} 
classical dynamics. In the case of a weak magnetic field, the effective 
kinetic operator corresponds to diffusion in the angle space, the diffusion 
constant being determined by the spectral resolution of the inhomogeneous
magnetic field. Applying these results to the quantum problem, we demonstrate 
that the low-lying modes of the field theory are related to the eigenmodes of 
the irreversible classical dynamics, and the higher modes are separated from 
the zero mode by a gap associated with the lowest density relaxation rate. 
As a consequence, we find that the long-time properties of the system are
characterised by universal Wigner-Dyson statistics. For a weak magnetic 
field, we obtain a description in terms of the quasi one-dimensional 
non-linear $\sigma$-model.

\end{abstract}

\section{Introduction}

Over recent years, the phenomenon of quantum chaos has been the subject of intense 
theoretical and experimental investigations~\cite{gutzwiller}. Of these, the 
most recent has been the development of a statistical field theory in which 
the quantum properties of classically chaotic systems are expressed through 
the modes of density relaxation specified by the classical evolution 
operator~\cite{muzykantskii}. In the present paper, we will apply 
this field theoretic procedure to investigate the quantum dynamics of a 
two-dimensional particle subject to a random magnetic field (RMF). Our choice is
motivated by two factors. Firstly, this problem represents one of the simplest
examples of a chaotic system. Secondly, it is of great practical interest,
and as such has already received a lot of attention in the 
literature~\cite{geim,smith,mancoff,ioffe,nagaosa,halperin,kalmeyer,hedegard,mirlin,khveshchenko,isichenko,zielinski,miller,aronov,altshuler}.
To put the present investigation in 
perspective, we will first describe the basis of the general field theoretic 
construction. Then, we will review the main results on the RMF problem, and
outline the strategy for the rest of the paper. 

The field theoretic approach to general chaotic quantum structures has been 
motivated by the success enjoyed by the statistical field theory of weakly 
disordered metallic conductors. More specifically, applied to a model in 
which time-reversal symmetry is broken, an average over realisations of a 
Gaussian $\delta$-correlated random impurity potential shows that the 
two-particle spectral properties of a weakly disordered Hamiltonian are 
described by a supersymmetric non-linear $\sigma$-model with the effective 
action~\cite{efetov}
\be \label{intr8}
S=-\frac{\pi\nu}{4}\int d{\bf r} {\rm str}\left[\hbar D\left(\nabla Q\right)^2 
+ 2 i s^+ \Delta\sigma_3^{\rm {\sc ar}} Q\right],
\ee
where $\nu=1/V\Delta$ denotes the density of states per unit volume $V$, $D$ 
defines the classical diffusion constant, and $s\Delta$ represents the 
symmetry breaking energy source with $\sigma_3^{\rm {\sc ar}}=
{\rm diag}(1,-1)_{\rm {\sc ar}}$. Here $Q({\bf r})$ represents a $4 \times 4$ 
supermatrix field obeying the non-linear constraint $Q^2({\bf r})=1$. (An
extensive review of this standard formalism can be found for example in 
Ref.~\cite{efetov}.)

The connection between the supersymmeric field theory and
physical coherence effects in disordered conductors can be understood
by analysing characteristic scales of the theory. 
The effective action~(\ref{intr8}) identifies two time scales, the diffusion
time $t_D=L^2/D$ and the Heisenberg time $t_H=\hbar/\Delta$. Their ratio 
defines the dimensionless conductance which in a good metal is large:
$g\equiv t_H/t_D\gg 1$. For energies $s\Delta \ll \hbar/t_D$, where 
quasi-classical dynamics is {\em ergodic}, the effective action is 
dominated by the zero spatial mode, $Q({\bf r})=Q_0$, independent of ${\bf r}$
\begin{eqnarray}
S[Q_0]=-i{\pi s^+\over 2}{\rm str}\left[\sigma_3^{\rm {\sc ar}} Q_0\right].
\end{eqnarray}
In this limit, spectral properties are universal and, in this case, 
coincide with those of random matrix ensembles of unitary symmetry.
Conversely, on energy scales $s\Delta\gg \hbar/t_D$, the effective action
is dominated by the global saddle-point $Q_{\rm sp}=\sigma_3^{\rm {\sc ar}}$.
An expansion in terms of the generators of the coset identifies the low-lying 
modes of density relaxation as diffusion modes. Interaction of these modes 
induces the well-known quantum weak localisation corrections.

Surprisingly, an analogous theory can be defined for systems which are 
non-integrable but not stochastic~\cite{muzykantskii}.
The key to the construction of an 
effective field theory of {\em individual} chaotic structures relies on the 
recognition that the spectrum itself provides a statistical ensemble~\cite{andreev}. 
By averaging over a wide interval of energy $N\Delta$ it is possible to develop 
a ``ballistic'' non-linear $\sigma$-model within the same general framework. 
In this case the effective action takes the {\em general} form
\be\label{intr22}
S[{\cal Q}]=-i{\pi\over 2}\int d\x_\parallel
{\rm str}\left [s^+\sigma_3^{\rm {\sc ar}}{\cal Q}+2i\hbar {\cal T}^{-1}
\hat{\cal L}{\cal T}\right].
\ee
where $\hat{\cal L}=\{H,\ \}$ denotes the classical evolution operator, and
the supermatrix field ${\cal Q}={\cal T}^{-1}\sigma_3^{\rm {\sc ar}}{\cal T}$ 
depends on the $2d-1$ phase space coordinates $\x_\parallel=({\bf r},
\p)_{2d-1}$ parametrising the constant energy shell $H({\bf r},\p)=E$. (Here
the coordinates are normalised such that $\int d\x_\parallel=1$.)
In the presence of a magnetic field, 
${\cal Q}$-matrices in (\ref{intr22}) correspond to the unitary ensemble and
are subject to additional symmetry constraints -- see Ref.~\cite{simons} for details.

While the zero-mode contribution to the action~(\ref{intr22}) reproduces
random matrix or Wigner-Dyson statistics~\cite{bohigas}, 
higher mode fluctuations establish 
non-universal quantum corrections. The utility of the field-theoretical 
approach beyond the universal regime was demonstrated in papers by Blanter 
{\em et al.}~\cite{blanter}, and Tripathi and Khmel'nitskii~\cite{tripathi} 
where the effective action (\ref{intr22}) was applied to the study of a 
quantum billiard with diffusive surface scattering, a system that exemplifies 
a ballistic system in the regime of strong chaos. However, the $\sigma$-model 
approach has so far failed to provide explicit results for truly ballistic 
systems. This can, in part, be attributed to the fact that the kinetic 
operator $\hat{\cal L}$, which enters the effective action, is antihermitian 
for any ballistic system, a signature of the reversible nature of classical 
dynamics. As a consequence, its eigenvalues lie on the imaginary axis, so 
that the time evolution of a distribution in the phase space does not exhibit 
relaxation into the uniform ergodic state. Moreover, for chaotic systems, 
$\hat{\cal L}$ is ill-defined in the following sense: Classical dynamics 
involves stretching along the unstable manifold and contraction along the 
stable one. Thus, any non-uniform initial distribution will evolve into a 
highly singular function. In terms of the field theory, it means that the 
functional integral suffers from ultraviolet divergencies. In order to 
understand this, let us notice that the action (\ref{intr22}) is only 
sensitive to the variations of the ${\cal Q}$-matrix along the classical
trajectories. Therefore, nothing prevents the ${\cal Q}$-matrix
from fluctuating in the transverse direction. It is these short-scale
fluctuations that ultimately lead to the divergence of the functional
integral~\cite{andreev}.

Several regularization procedures have been proposed to circumvent this 
problem~\cite{gaspard,ruelle2,nicolis,hasegawa}. 
One of the most natural ways to regularize 
the functional integral is to introduce an additional term into the effective 
action which suppresses the fluctuations of ${\cal Q}$ in the transverse 
direction. After performing the integration, one should take the regulator 
to zero. A surprising feature of the chaotic dynamics is that the limits
time-to-infinity and regulator-to-zero do not commute. In particular, the 
eigenvalues of the regularized kinetic operator in the limit 
regulator-to-zero remain complex, with finite real parts corresponding to 
relaxation rates into the equilibrium distribution. These complex 
eigenvalues, which are independent of the regularization procedure,
reflect intrinsic irreversible properties of the classical chaotic dynamics, 
and are known as Ruelle resonances or the Perron-Frobenius 
spectrum~\cite{ruelle1,pollicot}. 

One should stress that the direct regularization method described above is of
little practical use. This has been the main reason why all the attempts to 
apply the ballistic $\sigma$-model to real systems have been unsuccessful. In 
the present paper we propose a new general approach to the 
regularization problem, which allows one to construct an effective field 
theory for the low-lying part of the Perron-Frobenius spectrum, and is 
applicable for both classical and quantum systems. To describe the basis of
this approach, let us recall that the kinetic operator is defined in some 
Hilbert space, elements of which are smooth functions of the phase space 
coordinates. As was mentioned earlier, the eigenfunctions of $\hat{\cal L}$ 
are highly singular and lie outside the Hilbert space. However, by properly 
choosing a subspace of the full Hilbert space one can make $\hat{\cal L}$ a 
well-defined operator in the sense that its eigenfunctions belong to the same 
subspace. Obviously, this subspace should correspond to the physically 
relevant low-lying part of the Perron-Frobenius spectrum. As we will see 
later, the kinetic operator defined in such a way is irreversible. To 
summarize, instead of calculating the full spectrum of the regularized 
kinetic operator, one can instead obtain an effective operator that correctly 
describes the low-lying modes of the irreversible classical dynamics.

In order to choose the subspace in which the effective kinetic operator will
act, one should truncate the basis in the Hilbert space by eliminating degrees
of freedom irrelevant for the long-time evolution. Operationally, this is best 
done in the field-theoretical formulation. To illustrate our approach, let us 
consider the following formalism, wherein the Green 
function of the classical kinetic operator is represented as an integral over 
the superfield $\psi = (\psi_B, \psi_F)$,
\be \label{intr25}
\hat g (\omega)\equiv \frac{1}{i\omega - \hat{\cal L}}
=\int D\psi^* D\psi \, \psi_B \otimes \psi^*_B \exp\left [-\int d\x_ \parallel 
\psi^ \dag (\x_ \parallel)
( i \omega - \hat{\cal L} ) \psi (\x_ \parallel)\right ].
\ee
We then employ an RG-type approach and decompose $\psi$ in the following way,
\be \label{intr26}
\psi (\x_ \parallel) = \Psi (\x_ \parallel) + \chi (\x_ \parallel),
\ee
where $\Psi (\x_ \parallel)$ denotes ``slow'' fields, and 
$\chi (\x_\parallel)$ ``fast'' fields. Integrating over the ``fast'' fields 
$\chi$ one obtains an expression for the effective Green function $\hat G$ in 
the $\Psi$-subspace,
\be \label{intr27}
\hat g (\omega)
=\int D\Psi^* D\Psi \, \Psi_B \otimes \Psi^*_B \exp\left [-\int d\x_ \parallel 
\Psi^ \dag (\x_ \parallel)
( i \omega - \hat{\cal L}_{\rm eff} ) \Psi (\x_ \parallel)\right ],
\ee
containing a renormalized kinetic operator $\hat{\cal L}_{\rm eff}$. There is 
no universal recipe as to how one should identify the ``fast'' and ``slow'' 
fields. However, this can be done straightforwardly in the case of 
``weak non-integrability''\footnote{To eliminate possible confusion, we stress that the system is assumed to be fully ergodic throughout the paper; in other words, we are in the regime of hard chaos, and the KAM theorem is not applicable here.}, if the 
kinetic operator can be represented in the form
\be \label{intr28}
\hat{\cal L}=\hat{\cal L}_0 + \hat{\cal L}_{\rm pert},
\ee
where $\hat{\cal L}_0$ describes an integrable classical system, and
$\hat{\cal L}_{\rm pert}$ can be considered as a small perturbation.
In this case
one can, in principle, calculate the eigenfunctions and eigenfrequencies of the
unperturbed operator $\hat{\cal L}_0$. ``Fast'' fields are identified as
eigenfunctions of $\hat{\cal L}_0$ corresponding to high frequencies, while
``slow'' fields correspond to the low-frequency part of the spectrum of 
$\hat{\cal L}_0$.

The main purpose of the paper is to demonstrate how this general 
regularization procedure works by applying it to the problem of a particle 
confined to two-dimensions and propagating in a 
random magnetic field. The latter has attracted great interest in recent years. 
Firstly, there exist a number of experimental realizations where a RMF is
imposed on 2D electron gas. These include cases where the RMF is imposed by 
randomly pinned flux vortices in a type-II superconducting gate~\cite{geim},
by grains of a type-I superconductor~\cite{smith}, or by a demagnetized 
permanent magnet placed in the vicinity of the electron gas~\cite{mancoff}. 
Secondly, models of this type have been proposed within the gauge theory of 
high-$T_c$ superconductivity~\cite{ioffe,nagaosa}. Thirdly, the problem is 
thought to be relevant for the composite fermion theory of the quantum Hall 
effect near $\nu=1/2$~\cite{halperin,kalmeyer}.

The classical description of the transport properties of two-dimensional 
charged particles in a random magnetic field with {\em long-range} 
correlations can be given in terms of the distribution function 
$f({\bf r},\p)$, which obeys the Boltzmann equation
\be \label{intr29}
\frac{\partial f}{\partial t}+{\bf v}\frac{\partial f}{\partial {\bf r}}
+\frac{e}{c}[{\bf v},{\bf B}]\frac{\partial f}{\partial \p}=
-\frac{f-f_0}{\tau},
\ee
where the right-hand side describes relaxation to the uniform distribution
$f_0=\delta (v-v_F)$ due to impurity potential scattering. Assuming the 
magnetic field is weak, one can calculate the averaged Green function of 
Eq.~(\ref{intr29}) using a perturbation theory. Within the Born approximation 
the renormalized value of the transport scattering time $\tau_{tr}$ can be 
found~\cite{hedegard}. In particular, in the absence of the potential 
scattering ($\tau\to\infty$), one obtains for the transport time~\cite{mirlin}
\be \label{intr30}
\frac{1}{\tau_{tr}}=\frac{1}{mp_F}\int_0^\infty dr \langle B({\bf 0}) 
B({\bf r}) \rangle,
\ee
where the angle brackets denote the ensemble average. 

In the opposite case of a strong magnetic field, where the Born approximation 
is not applicable, a number of other approaches have been developed for 
solving the Boltzmann equation, including the ``Eikonal'' 
approximation~\cite{khveshchenko}, and the percolation theory~\cite{mirlin}. 
Another possible approach is to use known results from the time-independent 
diffusion-advection or Passive scalar equation
\be \label{intr31}
\frac{\partial n}{\partial t} = D \nabla^2 n + {\bf u} \cdot \nabla n,
\ee
where ${\bf u}$ is a random velocity field. Indeed, starting with 
Eq.~(\ref{intr29}) and assuming a rapid relaxation in the momentum space, one 
obtains an effective equation for the particle density $n({\bf r})=\int 
d\p f({\bf r},\p)$, which has the form of Eq.~(\ref{intr31}). It can be 
interpreted as the advection of guiding centers (or ``van Alfven drift'' of 
cyclotron orbits) due to the random component of the magnetic field. The 
diffusion-advection problem has been extensively studied because of its
importance in fluid dynamics and plasma physics~\cite{isichenko}. Known results
suggest a non-trivial scaling behaviour of the conductivity $\sigma_{xx}$, 
which has been confirmed in recent experiments~\cite{zielinski}. It is also 
interesting to note that an analogy exists between the diffusion-advection 
problem and the localization problem for a quantum particle in a random 
vector potential~\cite{miller}.

The quantum transport properties of a particle subject to a random magnetic 
field have been studied by Aronov, Mirlin, and W\"olfle~\cite{aronov}. 
Applying a Gaussian $\delta$-correlated 
distribution for the magnetic field, spectral and transport properties are 
shown to be described by a 
diffusive supersymmetric non-linear $\sigma$-model of unitary symmetry.
Although the relaxation rate which enters the average 
single-particle Green function is divergent if calculated perturbatively, the 
transport relaxation time, which is a characteristic of the average 
two-particle Green function, turns out to be finite. The question of finding 
a physically meaningful definition of the single-particle relaxation time has 
been discussed by Altshuler {\em et al.}~\cite{altshuler}.

In contrast to the above work by Aronov {\em et al.}, we want to focus on the case
of a weak slowly-varying magnetic field. It is convenient for our purposes to
divide the rest of the paper into two sections. In the first section, we 
deal with the classical problem. The procedure of separating slow and fast 
fields, outlined above, is carried out in detail, and the effective classical 
kinetic operator is evaluated. The latter is shown to contain a term 
describing diffusion in the angle space. We calculate the corresponding 
``diffusion constant'' (transport time), which is determined by the power 
spectrum of the magnetic field. The angle diffusion term is responsible for 
the relaxation of the momentum-dependent degrees of freedom, leading to the 
conventional diffusive dynamics at large scales. 

In the second section, we consider the quantum problem, and apply the 
quasi-classical field theoretic description~(\ref{intr22}). We show that,
cast in the form of a field theory, the classical and quantum problems are 
closely related. One can therefore separate slow and fast degrees of freedom 
in Eq.~(\ref{intr22}) in almost the same way as one does for the classical 
problem. The only complication is the somewhat more complex structure of the 
integration manifold. Taking this into account, and using results from the 
first section, we arrive at the following conclusions: At short scales (small 
system size), spectral properties are described by a quasi-one-dimensional 
$\sigma$-model. At intermediate scales, we obtain a ballistic $\sigma$-model, 
where the traditional collision integral is replaced with diffusion in the 
angle space. At very large scales, we recover the conventional diffusive 
$\sigma$-model. In this way, we demonstrate that a description in terms of 
the diffusive $\sigma$-model holds not only for systems with 
$\delta$-correlated magnetic fields dominated by quantum 
scattering~\cite{aronov}, but also for semi-classical motion in a long-range 
magnetic field. In other words, although the semi-classical description is 
formally valid, quantum corrections are essential, and the transport at large 
scales is dominated by localization effects. Finally, it is important to 
emphasize that, in contrast to the existing literature on this problem, we 
do not make use of the notion of an ensemble average, so that our results 
are valid for an {\em individual} system with a given configuration of the 
magnetic field. 

\section{Classical Dynamics}

As mentioned above, as a precursor to the analysis of the quantum problem, we
begin by considering the dynamics of a {\em classical} particle in 
two-dimensions subject to an inhomogeneous perpendicular magnetic field.
Indeed, provided that $\hbar/mv\ll a$, where $m$ is the mass of the particle, 
$v$ is its velocity, and $a$ is the typical length scale at which the 
magnetic field fluctuates, the motion of the quantum particle can be described 
semi-classically.
Since energy is conserved, $|{\bf v}|={\rm const.}$, the motion is defined
by the set of phase space variables $(x,y,\phi)$, which obey the following 
equations
\begin{eqnarray}
\dot x  = v\cos\phi,\qquad \dot y  = v\sin\phi,\qquad \dot \phi = \Omega(x,y), 
\end{eqnarray}
where $\Omega =eB/mc$ represents the cyclotron frequency. We will assume that 
$\Omega(x,y)$ is a random function of the spatial coordinates with zero 
average. We will further assume that the magnetic field is weak: $\Omega a \ll v$. 

As the particle moves through areas of different magnetic field, its 
trajectory is slightly deflected to one side or another depending on the sign 
of the magnetic field. From the point of view of the particle, the magnetic 
field $\Omega (x(t),y(t))$ is a random function of time. As such, one can 
obtain a good qualitative description of the particle's motion by taking this 
function to be $\delta$-correlated Gaussian noise:
\begin{eqnarray}
\dot \phi & = & \Omega(t), \nonumber\\
\bra \Omega(t)\Omega(t')\ket & = & \frac{2}{\tau_{tr}}\delta(t-t').
\label{magn4}
\end{eqnarray}
From dimensional considerations it is clear that $1/\tau_{tr} \sim \Omega^2 
a/v$. From Eq.(\ref{magn4}) one obtains
\be \label{magn5}
\bra \phi^2(t)\ket =\int_0^t \int_0^t dt' dt''
\bra \Omega (t')\Omega(t'') \ket =\frac{2}{\tau_{tr}}t.
\ee
Such time dependence of $\bra \phi^2 \ket$ corresponds to diffusion in the 
angle space, with the diffusion coefficient $1/\tau_{tr}$. If $t\lesssim a/v$, 
the motion is ballistic, and our description fails. At times $a/v \ll t \ll 
\tau_{tr}$ (the existence of this intermediate regime is ensured by the 
condition of weakness of the magnetic field $\Omega a \ll v$) one has 
$\bra \phi^2 \ket \ll 1$, so that the particle's trajectory in real space 
remains an approximately straight line. Finally, at times $t \gtrsim 
\tau_{tr}$, angle diffusion results in the uniform distribution of angles. 
Physically this means that the direction of movement becomes random, which 
leads to diffusion in real space (diffusive regime). Indeed, one has
\be \label{magn6}
\bra x^2 (t)\ket =v^2 \int_0^t \int_0^t dt' dt''
\bra \cos\phi(t')\cos\phi(t'')\ket \approx 2Dt,
\ee
where
\be \label{magn7}
D=\frac{v^2 \tau_{tr}}{2}
\ee
represents the diffusion coefficient in real space.

This description offers a clear physical picture of what is going on in the 
system. However, it has some drawbacks: Firstly, since $\delta$-correlation 
contradicts the existence of a finite correlation length $a$ of the magnetic 
field, it is not self-consistent. Secondly, $a$ and $\Omega$ themselves are not 
well-defined quantities. Afterall, this is
just a phenomenological description. We would like to obtain a quantitative
description that would relate the diffusion coefficient to specific spectral
properties of the random function $\Omega (x,y)$. In order to do this we will 
make use of an alternative representation of the problem in the form of the 
Boltzmann equation. This representation has the advantage of being linear, 
allowing us to apply a field-theoretical approach.

We start by recalling some basic facts about the Boltzmann equation: Any 
classical system obeying Hamilton's equations of motion can be alternatively 
described in terms of the distribution function $f(\x)$, which obeys the 
following partial differential equation
\be \label{magn9}
\frac{\partial f}{\partial t}+\hat{\cal L}f =0,
\ee
where the kinetic operator $\hat{\cal L}$ describes an incompressible flow in
the phase space,
\be \label{magn15}
\hat{\cal L}=\dot\x\cdot\frac{\partial}{\partial \x}, \qquad \partial\cdot
\dot\x =0.
\ee

The incompressibility of the flow has a number of important implications, 
including the conservation of particle number, and the antihermiticity of
$\hat{\cal L}$. In particular, for two arbitrary states $|\psi\rangle$ and 
$|\chi\rangle$, the following relation holds:
\be \label{magn18}
\langle\psi|\hat{\cal L}|\chi\rangle^*= 
-\langle\chi |\hat{\cal L}|\psi\rangle.
\ee
Consequently, the eigenvalues of $\hat{\cal L}$ are purely imaginary numbers.
In the introduction we discussed the reasons why the spectrum of the kinetic
operator for a chaotic system is not well-defined, and came to the conclusion
that the kinetic operator should be regularized. In the simplest case of a free
particle, this can be done by introducing a term which accounts for the
relaxation of the fast modes into the right-hand side of Eq.~(\ref{magn9}):
\be \label{magn19}
\frac{\partial f}{\partial t}+{\bf v} \cdot \frac{\partial f}{\partial {\bf r}}=
u\,\partial^2 f,\qquad u > 0.
\ee
Substituting $f=f_{\k,\omega}\exp ( i\k\cdot{\bf r}-i\omega t )$ into 
Eq.~(\ref{magn19}) and taking the limit $u \rightarrow 0$, one finds that the 
eigenvalues $i\omega (\k)$ of $\hat{\cal L}$ acquire an infinitesimal 
positive real part:
\be \label{magn21}
\omega({\k})={\bf v}\cdot\k -i 0.
\ee
As suggested in the introduction, the eigenvalues of a generic 
Perron-Frobenius operator have {\em finite} positive real parts.

For the problem at hand, it is convenient to introduce the parametrization
\be \label{magn12}
{\bf v}=v\n(\phi),\qquad \n(\phi)=(\cos\phi,\sin\phi).
\ee
The kinetic energy of a particle in the magnetic field is conserved, so that 
the phase space $\x=({\bf r},\p)$ is effectively reduced to the energy shell
$\x_\parallel =({\bf r},\phi)$, and the kinetic operator takes the form
\be \label{magn14}
\hat{\cal L}=v\n(\phi)\cdot\frac{\partial}{\partial {\bf r}}
+\Omega({\bf r})\frac{\partial}{\partial \phi}.
\ee

Since $\hat{\cal L}$ is a linear operator, Eq.~(\ref{magn9}) can be solved. 
The general solution can be written in the form
\be \label{magn22}
f(\x,t)=\hat U (t) f(\x,0),
\ee  
where $\hat U (t) = \exp (-\hat{\cal L}t)$ represents the classical evolution 
operator. It is often more convenient to work with its Fourier transform
\be \label{magn24}
\hat g (\omega) =-\int_{0}^{\infty}dt e^{i\omega t}\hat U (t)=
\frac{1}{i\omega -\hat{\cal L}}.
\ee

If we had an ensemble of different configurations of the magnetic field,
information about transport properties could be extracted from the 
ensemble averaged distribution function $\bra f(\x,t)\ket$. As follows from 
Eqs.~(\ref{magn22}) and (\ref{magn24}), it would be sufficient to calculate 
the ensemble averaged evolution operator $\langle \hat U (t)\rangle$ or the 
Green function $\langle \hat g (\omega) \rangle$. One could then introduce 
the effective ``ensemble averaged'' kinetic operator $\hat{\cal L}_{\rm eff}$ 
in the following manner:
\be \label{magn25}
\langle \hat U (t) \rangle =\exp (-\hat{\cal L}_{\rm eff}t).
\ee
Since $\hat{\cal L}$ is antihermitian, $\hat U (t)$ is a unitary operator.
The ensemble averaged unitary operator, however, is not necessarily unitary,
which means that $\hat{\cal L}_{\rm eff}$ is {\em not} antihermitian, so that 
the spectrum of $\hat{\cal L}_{\rm eff}$ has a real part. Completely 
reversible dynamics becomes irreversible after the averaging.

One can, in principle, derive the spectrum of relaxation modes by 
calculating $\langle\hat g (\omega)\rangle$. But, since we are dealing with an
individual system, not with an ensemble of systems, we are bound to take a
different route. Following the ideas explained in the introduction, we 
therefore represent the classical Green function (\ref{magn24}) as a field 
integral,
\be \label{magn29}
\hat g(\omega)=\int D\psi^* D\psi \, \psi_{B} \otimes \psi^*_{B}
\exp \{-S[\psi^*,\psi]\},
\ee
where 
\be \label{magn30}
S=\int d{\bf r} d\phi \, \psi^\dag \left(i\omega- \hat {\cal L}\right)\psi
\ee
denotes the effective action, and $\psi({\bf r},\phi) = (\psi_{B},\, \psi_{F})$ 
represents a two-component supervector field, with indices B and F 
corresponding to bosonic and fermionic components. To make the functional
integral (\ref{magn29}) formally convergent, one usually puts an infinitesimal
regulator into the action (\ref{magn30}). This is essentially the same regulator
as in Eq.~(\ref{magn21}), and it will play an important role in the following
analysis.  

The first term of the operator (\ref{magn14}) describes the evolution of a free 
particle. Note that any function $\Psi(\phi)$ which depends on $\phi$ but 
not on ${\bf r}$ is necessarily a zero-mode of $\hat{\cal L}_{\rm free}
\equiv v \n (\phi)\cdot \partial/\partial {\bf r}$ (i.e. 
$\hat{\cal L}_{\rm free}\Psi(\phi)=0$). This implies that the angle 
distribution (in $\phi$-space) remains constant in time. Switching on a weak 
magnetic field, the angle distribution starts changing slowly. The evolution 
in $\phi$-space is described by some effective operator which, following on 
from our naive physical picture, must have a diffusive form with some 
diffusion coefficient $1/\tau_{tr}$. As we will see later, $1/\tau_{tr}$ is 
determined by spectral properties of $\Omega({\bf r})$.

This observation suggests a separation into slow and fast degrees of freedom 
from which the fast degrees of freedom can be integrated away. Since the 
particle's velocity is high and the magnetic field is weak, the spatial 
coordinate ${\bf r}(t)$ of the particle changes rapidly, while
$\phi (t)$ changes slowly. The same statements are true for the 
distributions in the ${\bf r}$-space and $\phi$-space respectively.
In terms of the action, one could say that spatially-dependent modes are fast,
and spatially-independent modes are slow. The corresponding timescales are
$\tau_c = L/v$ and $\tau_{tr}$, where $L$ is the size of the system.
One can always choose $L$ such that $\tau_c \ll \tau_{tr}$, and at the same time
$L \gg a$, the consistency of the two inequalities being ensured by the weakness
of the magnetic field. (This is true only if the correlation length $a$ of the 
magnetic field is finite -- see below.) In this case, 
one can separate the modes with $\k =0$ and perform
the integration over the rest of the modes. By doing so, one obtains an 
effective action describing the evolution in the $\phi$-space. For larger systems,
a spatial dependence becomes important. The latter can be accounted for by
dividing the system into smaller sub-systems and repeating the above procedure
for each sub-system. Finally, in the case of very large systems,
$\tau_c \gg \tau_{tr}$, the effective action is dominated by spatial diffusion modes. 

Beginning with small systems, we adopt the representation
\be \label{magn32}
\psi({\bf r},\phi)=\frac{1}{\sqrt{V}}\Psi(\phi)+\frac{1}{\sqrt{V}}
\sum_{\k \not = 0} \chi_\k (\phi) e^{i\k\cdot{\bf r}}, \qquad 
\Omega({\bf r})=\frac{1}{\sqrt{V}}\sum_{\k \not = 0} \Omega_\k e^{i\k\cdot
{\bf r}},
\ee
where $\chi_\k (\phi)$ are small fluctuating fields, and $V$ denotes the 
volume of the system. Substituting Eq.~(\ref{magn32}) into (\ref{magn30}), we
obtain
\be \label{magn35}
S=\int d\phi \Bigg\{ i\omega \Psi^\dag \Psi-\sum_{\k\not=0}\left [ iv
\k\cdot\n(\phi)\chi^\dag_\k \chi_\k + \frac{1}{\sqrt{V}}\chi^\dag_\k 
\Omega_\k\frac{\partial \Psi}{\partial \phi}-\frac{1}{\sqrt{V}}\Omega^*_\k 
\frac{\partial \Psi^\dag}{\partial \phi}\chi_\k \right ]
\Bigg\},
\ee
where we have kept terms of up to second order. Performing the Gaussian
integration over $\chi_\k (\phi)$, we obtain
\begin{eqnarray} 
\hat g(\omega) = \frac{1}{V} \int D\Psi^* D\Psi \, \Psi_{B} \otimes 
\Psi^*_{B}\exp \{-S_{\rm eff}[\Psi^*,\Psi]\},
\label{func_int}
\end{eqnarray}
where
\begin{eqnarray}
S_{\rm eff} = \int d\phi \left\{ i\omega \Psi^\dag \Psi-\frac{1}{\tau_{tr}}
\frac{\partial \Psi^\dag}{\partial \phi}\frac{\partial \Psi}{\partial \phi}
\right\} \label{magn38}
\end{eqnarray}
denotes the effective action, and 
\be \label{magn39}
\frac{1}{\tau_{tr}}=\frac{1}{V}\sum_\k \frac{|\Omega_\k|^2}{i\v\cdot\k}.
\ee
(Note that consistency requires only the zeroth-order term to be kept in the 
pre-exponential.)
Replacing the summation in Eq.~(\ref{magn39}) by integration,
and introducing the infinitesimal regulator according to Eq.~(\ref{magn21}), 
one finds
\be \label{magn40}
\frac{1}{\tau_{tr}}=-i\int\frac{|\Omega_\k|^2 d\k}{\v\cdot\k - i 0}=
\frac{2\pi}{v}\int_0^\infty dk |\Omega_k|^2,
\ee
assuming that the spectrum of $\Omega({\bf r})$ is isotropic, i.e. depends 
only on $|\k|$. The effective action (\ref{magn38}) corresponds to the 
kinetic operator of the form
\be \label{magn42}
\hat{\cal L}_{\rm eff}=-\frac{1}{\tau_{tr}}\frac{\partial^2}{\partial\phi^2}.
\ee
The distribution function in $\phi$-space $\tilde f(\phi)=\int d{\bf r} 
f({\bf r},\phi)$ obeys the diffusion equation
\begin{equation} \label{magn43}
\frac{\partial \tilde f}{\partial t}=\frac{1}{\tau_{tr}}
\frac{\partial^2 \tilde f}{\partial \phi^2}.
\end{equation}
Note that the functional integral (\ref{func_int}) is convergent, since the
eigenvalues of $\hat{\cal L}_{\rm eff}$ are non-negative. In other words,
$\hat{\cal L}_{\rm eff}$ plays the role of a {\em finite} regulator in the
field-theoretic representation. 

In conclusion, having started with a completely reversible kinetic operator 
with an imaginary spectrum, we ended up with an effective irreversible operator
describing a relaxation into the uniform distribution. It is important to 
emphasize that we did not introduce any collision integral. On the formal 
level, irreversibility comes from the infinitesimal imaginary part of the 
spectrum (\ref{magn21}). One can draw an analogy between this problem and
the problem of Landau damping in a plasma~\cite{lifshitz}. 
For the latter, the infinitesimal
imaginary part of the spectrum (\ref{magn21}) results in a finite damping rate
of the {\em collisionless} plasma. One could say that the initial infinitesimal
relaxation rate (which one has to introduce simply to choose the direction of
the time evolution) is made finite by the regularization procedure, which for 
the problem at hand consists of truncating the phase space by eliminating 
large-$\k$ modes. This is quite a general situation: Many systems which 
exhibit irreversible properties such as increase of entropy and relaxation to 
equilibrium, are actually described by reversible classical mechanics. This 
phenomenon can be illustrated by the simple example of a gas of classical 
particles. A proper description must be in terms of the many-particle 
distribution function $f(\x_1, \dots ,\x_N)$. The corresponding kinetic 
operator $\hat{\cal L}(\x_1, \dots ,\x_N)$ is, of course, fully reversible, 
that is having a purely imaginary spectrum. In order to obtain a description 
in terms of the two-particle distribution function $\tilde f(\x, \x')$ one has 
to truncate the phase space. As a result, the effective operator 
$\hat{\cal L}_{\rm eff}(\x, \x')$ contains a collision integral, which 
describes a relaxation into the Maxwell distribution. As shown by this 
example, there is nothing surprising in the fact that classical dynamics have 
irreversible properties.

For the diffusion constant (\ref{magn40}) to be finite, there has to be
a finite correlation length of the magnetic field. For a $\delta$-correlated
magnetic field one would have $|\Omega_k|^2={\rm const.}$, and the integral 
(\ref{magn40}) for $1/\tau_{tr}$ would diverge at large $k$, making the separation
of modes illegitimate. We assume that the case $a=0$ corresponds to a different
physical regime which can not be described by a diffusive approximation.

Equation (\ref{magn42}) is only valid if the distribution in the real space
becomes uniform at times much shorter than the diffusion time in the angle 
space. This is the case if the size of the system $L$ is not too large.
In the opposite case of a very large system, $L/v \gg \tau_{tr}$, one
has to follow the spirit of the renormalisation group approach and divide
the large system into smaller sub-systems. In each sub-system, one can
separate the modes with $\k=0$ and repeat previous calculations, integrating out
the higher modes. The size of a sub-system should be small enough, but still much
larger than the correlation length $a$ of the magnetic field. Since each sub-system 
can be considered as independent, one has to 
introduce a coordinate ${\bf R}$ that labels the sub-systems:

\be \label{magn45}
\psi({\bf r},\phi)=\frac{1}{\sqrt{V}}\Psi(\R,\phi)+\frac{1}{\sqrt{V}}
\sum_{\k \not = 0} \chi_\k (\R,\phi) e^{i\k\cdot{\bf r}},
\ee
where $V$ stands for the volume of a sub-system.
This
representation effectively separates the degrees of freedom with small and large
momenta. Substituting (\ref{magn45}) into (\ref{magn30}), keeping the terms of up 
to the second order, and using the fact that the main contribution to the 
magnetic field comes from large momenta, we obtain:

\begin{eqnarray} \label{magn46}
S=\int d\R d\phi \left\{i\omega \Psi^\dag \Psi -
\sum_{\k\not=0}\left [ iv\k\cdot\n(\phi)\chi^\dag_\k \chi_\k + 
\frac{1}{\sqrt{V}}\chi^\dag_\k \Omega_\k
\frac{\partial \Psi}{\partial \phi}-\frac{1}{\sqrt{V}}\Omega^*_\k 
\frac{\partial \Psi^\dag}{\partial \phi}\chi_\k \right ] \right.\\
\left. -\Psi^\dag v\n(\phi)\cdot\frac{\partial\Psi}{\partial \R}\right\}. \nonumber
\end{eqnarray}
Integration over $\chi_\k (\R,\phi)$ and $\chi^*_\k (\R,\phi)$ yields
\begin{equation} \label{magn47}
S_{\rm eff} = \int d{\bf R}d\phi \left \{ i\omega \Psi^\dag \Psi -
\frac{1}{\tau_{tr}}\frac{\partial \Psi^\dag}{\partial \phi}
\frac{\partial \Psi}{\partial \phi} 
-\Psi^\dag v\n(\phi)\cdot\frac{\partial \Psi}{\partial {\bf R}}
\right\},
\end{equation}
which in turn implies a kinetic operator of the form
\be \label{magn48}
\hat{\cal L}_{\rm eff}=-\frac{1}{\tau_{tr}}\frac{\partial^2}{\partial \phi^2}
+v\n(\phi)\cdot\frac{\partial}{\partial {\bf R}},
\ee
with the same transport relaxation time (\ref{magn40}).

We are most interested in the spectrum of the effective kinetic operator, 
which, as explained in the introduction, coincides with the low-lying part of 
the Perron-Frobenius spectrum. Representing an eigenfunction in the form
\be \label{magn50}
\Psi({\bf R},\phi)=\psi(\phi)e^{i\q\cdot{\bf R}},
\ee
we obtain Mathieu's equation, which is formally equivalent to the
Schr\"odinger equation for a quantum rotator in a uniform electric field,
\be \label{magn51}
-\frac{1}{\tau_{tr}}\psi^{\prime\prime} +ivq\cos\phi\, \psi =i\omega\psi.
\ee
Equation (\ref{magn51}) can not be solved analytically. However, if one is 
interested in the low-lying excitations, it can be recast in the form 
suitable for a perturbative treatment:
\be \label{magn55}
(\hat H_0 +\hat V)\psi = i\omega\psi,\qquad \hat H_0 =-\frac{1}{\tau_{tr}}
\frac{d^2}{d\phi^2},\qquad \hat V =ivq\cos\phi.
\ee
The unperturbed eigenfunctions and eigenvalues are
\be \label{magn57}
\psi^{(0)}_m=\frac{1}{\sqrt{2\pi}}e^{im\phi}, \quad
i\omega^{(0)}_m=\frac{1}{\tau_{tr}} m^2, \quad
m=0,\pm 1,\pm 2, \dots .
\ee
Non-zero transition matrix elements exist only between neighbouring states,
\be \label{magn59}
V_{m-1, m} =
V_{m, m-1} = \frac{ivq}{2}.
\ee
Therefore, the first order corrections are zero. Calculating the second-order 
correction to the eigenvalue of the ground state ($m=0$), we find
\begin{equation} \label{magn61}
i\omega(\q)=i\omega^{(2)}_0  = 
\sum_{m (\not =0)}\frac{V_{0,m}V_{m,0}}{i\omega^{(0)}_0-i\omega^{(0)}_{m}}=
Dq^2.
\end{equation}
The low-lying modes of the effective operator (\ref{magn48}) are nothing but 
diffusion modes in real space. The real space diffusion coefficient $D$ is 
related to the transport time $\tau_{tr}$ by the usual equation (\ref{magn7}).

So far, we have shown that the effective Perron-Frobenius operator for the 
problem at hand contains a term that describes relaxation in the angle space. 
The effect of this term at large scales is to make the dynamics diffusive. 
Under the hypothesis of ergodicity (self-averaging), our formula 
(\ref{magn40}) for the inverse scattering time matches the result obtained in 
the Born approximation after ensemble-averaging, Eq.~(\ref{intr30}). The 
advantage of the field-theoretical approach, however, is that it can be 
readily generalized to the quantum case, which is the subject of the next 
section.

\section{Quantum Dynamics}

While the short-time dynamics of a particle in a slowly-varying magnetic
field can for the most part be treated classically, the long-time dynamics 
are influenced strongly by quantum coherence effects which have manifestations
both in localisation and spectral properties. To study the influence of these
coherence effects on the quantum dynamics we will employ the statistical 
field theory defined in the introduction and applied to the quantum 
Hamiltonian
\be \label{magn65}
\hat H = \frac{1}{2m}\left ( \hat \p -\frac{e}{c}{\bf A}\right )^2,
\ee
where ${\bf B}=\partial\times{\bf A}$.

The action, defined by Eq.~(\ref{intr22}), retains all 
the advantages of a simple classical description in the language of the 
kinetic equation, at the same time accounting for quantum interference 
effects. However, to apply the quasi-classical theory, we must introduce an
appropriate regularization procedure. In doing so, we will take a lesson 
from the approach employed in the previous section.

Let us first establish a connection between the quantum and the classical 
field theories, by making an expansion of the action~(\ref{intr22}) around 
the saddle point ${\cal Q}=\sigma_3^{\rm {\sc ar}}$, which is valid at large 
frequencies $\omega = s\Delta/\hbar$. (In doing so, we make use of symmetry 
properties of ${\cal T}$'s -- see Ref.~\cite{simons}.) 
Applying the parametrization
\be \label{magn72}
{\cal T}= \openone + i \pmatrix{0 & \bar{\cal B} \cr {\cal B} & 0}_{\rm {\sc ar}},
\ee
expanding to the second order in $\bar{\cal B}$ and ${\cal B}$, and using the
antihermiticity of $\hat{\cal L}$, we recover the classical action
\be \label{magn74}
S[\bar{\cal B},{\cal B}]=\hbar\nu\int d{\bf r} d\phi {\rm str}
\left [ \bar{\cal B}\left(i\omega-\hat{\cal L}\right){\cal B} \right ].
\ee
At smaller frequencies, higher-order terms in the expansion generate quantum 
corrections. 

Assuming that the time scale at which the motion in real space 
becomes ergodic is set by the ballistic transport time across the system
$\tau_c=L/v$, we conclude that it is the spatially-dependent degrees of 
freedom that should be integrated out, in the same way as for the classical 
problem. Quantum effects will come from the interaction between the low-lying 
relaxation modes of the effective kinetic operator, which is induced by the 
non-linearity of the manifold of ${\cal Q}$-matrices.

In order to separate slow and fast degrees of freedom in the action 
(\ref{intr22}), instead of the linear decomposition (\ref{intr26}) one uses 
the following non-linear decomposition,
\begin{equation}
{\cal T}(\x_\parallel)= \tilde T (\x_\parallel) T (\x_\parallel),
\end{equation} 
where $\tilde T(\x_\parallel)$ stands for fast fields, $T(\x_\parallel)$ 
-- for slow fields. It ensures that the ${\cal Q}$-matrix always stays on 
the manifold ${\cal Q}^2 = \openone$. As explained above, fast fields are 
spatially-dependent, while slow fields depend only on angle $\phi$. Since 
$\tilde T(\x_\parallel)$ fluctuates only weakly around $\tilde T(\x_\parallel)
=\openone$, we arrive at the representation
\begin{equation} \label{magn75}
{\cal T}({\bf r},\phi)  =\left [ \openone + \frac{i}{\sqrt{V}} 
\sum_{\k \not = 0} \pmatrix{0 & \bar B_{-\k}(\phi) \cr B_\k (\phi) & 0 
\cr}_{\rm {\sc ar}}e^{i\k\cdot{\bf r}} \right ]T(\phi), 
\end{equation}
where $T(\phi)$ describes spatially-uniform modes, $B_\k(\phi)$ and 
$\bar B_\k(\phi)$ are small fluctuating fields describing the modes with 
non-zero momenta. Supermatrices $T(\phi)$ and $T^{-1}(\phi)$ parametrize the 
supermatrix $Q(\phi)$,
\be \label{magn76}
Q=T^{-1}\sigma_3^{\rm {\sc ar}} T,
\ee
so that $Q(\phi)$ obeys the non-linear constraint $Q^2=\openone$. It is also 
convenient to introduce the notation
\be \label{magn79}
M = T \partial_\phi Q T^{-1}.
\ee
Using the relation $[Q,\partial_\phi Q]_+ =0$, one can prove the identity
\be \label{magn85}
2 {\rm str}(M_{\rm {\sc ar}} M_{\rm {\sc ra}})={\rm str} ( \partial_\phi Q )^2.
\ee

Now we are prepared to derive an effective field theory for the problem at 
hand. Substituting the representation (\ref{magn75}) into the action 
(\ref{intr22}) with the kinetic operator (\ref{magn14}), and making an 
expansion in $B$ and $\bar B$, we obtain
\begin{eqnarray}
S=-\hbar\nu\int d\phi \sum_{\k \not =0}{\rm str}\left [ 
iv\k\cdot\n(\phi)\bar B_\k B_\k +\frac{i}{2}\bar B_\k \Omega_\k M_{\rm {\sc ar}} +
\frac{i}{2}\Omega_\k^* M_{\rm {\sc ra}}B_\k
\right ] \\
-\frac{i\hbar\omega}{4\Delta}\int d\phi {\rm str}
\sigma_3^{\rm {\sc ar}} Q. \nonumber
\end{eqnarray}
Integration over the fast degrees of freedom yields the effective action
\be \label{magn86}
S_{\rm eff}=-\frac{\hbar}{8\Delta}\int d\phi {\rm str}\left [ 
\frac{1}{\tau_{tr}} ( \partial_\phi Q )^2 + 2i \omega \sigma_3^{\rm {\sc ar}} 
Q \right ],
\ee
which has the form of a one-dimensional diffusive $\sigma$-model, the 
diffusion taking place in the angle space. The transport time is given by the 
same formula (\ref{magn40}) as for the classical problem. To arrive at 
Eq.~(\ref{magn86}), we have used the identity (\ref{magn85}).

If $\omega\tau_{tr} \gg 1$, the main contribution to the functional integral 
comes from around the saddle point $Q=\sigma_3^{\rm {\sc ar}}$. In the 
opposite case, $\omega\tau_{tr} \ll 1$, one must take into account the whole 
supermatrix manifold. Spectral correlation functions are then 
determined by universal Wigner-Dyson statistics. Corrections to this universal 
behaviour can 
be calculated by performing an expansion in $1/g$ analogous to that employed 
by Kravtsov and Mirlin~\cite{kravtsov}, where
\be \label{magn89}
g=\frac{\hbar}{\Delta\tau_{tr}}
\ee
plays the role of dimensionless conductance.

At larger scales the low-lying modes of the Perron-Frobenius operator acquire 
a spatial dependence. This can be accounted for by using another 
representation, instead of (\ref{magn75}), that separates degrees of freedom 
with small and large momenta (compare with the previous section).
The effective action takes the form of the ballistic $\sigma$-model,
\be \label{magn90}
S_{\rm eff}=\frac{\hbar\nu}{4}\int d{\bf r} d\phi {\rm str}\left [
2 T^{-1} \sigma_3^{\rm {\sc ar}} v \n(\phi)\cdot\nabla T -\frac{1}{2\tau_{tr}} 
(\partial_\phi Q)^2 + i \omega \sigma_3^{\rm {\sc ar}} Q \right ].
\ee

Note that a similar type of action was proposed by Muzykantskii and 
Khmelnitskii~\cite{muzykantskii}, where a collision integral was introduced 
as a phenomenological term to describe quantum scattering by a 
$\delta$-correlated impurity potential. The second term in the action 
(\ref{magn90}) describes the diffusion in the angle space and effectively 
plays the role of a collision integral in the sense that it is responsible 
for relaxation of the momentum-dependent degrees of freedom. However, its 
origin in our case is quite different, since it appears as a result of the 
regularization of the field theory. 

Finally, one can establish a relation between the ballistic $\sigma$-model 
(\ref{magn90}) and the conventional diffusive $\sigma$-model using the 
procedure similar to that of Ref.~\cite{muzykantskii}. Anticipating a rapid 
relaxation in the momentum space, and a slow variation of the spatial modes, 
one can integrate out the momentum-dependent degrees of freedom to obtain the 
action (\ref{intr8}) with the diffusion constant (\ref{magn7}). Therefore, 
the transport properties of the system at very large scales are dominated by 
the localization effects.

To summarise, in this paper we applied the recently-developed field theory of
quantum chaos to the problem of a particle propagating in a non-uniform magnetic
field. We proposed a new approach to the regularisation of the ballistic 
$\sigma$-model (\ref{intr22}), and obtained an effective description of  
spectral and transport properties of the quantum system in terms of the
low-lying modes of the {\em irreversible} dynamics of its classical counterpart.
We argued that this method is quite general and can be applied to a variety
of chaotic systems. 

{\em Acknowledgements:} We would like to thank Kirill Samokhin for stimulating 
discussions.

\end{document}